**Haptic guidance improves the visuo-manual tracking of trajectories**

Jérémy Bluteau[123], Sabine Coquillart[1], Yohan Payan[2] and Edouard Gentaz[3]*

1. Laboratoire d'Informatique de Grenoble (LIG), Institut National de la Recherche en Informatique et Automatique (INRIA) Grenoble-Rhône-Alpes, France

2. Laboratoire des Techniques de l'Ingénierie Médicale et de la Complexité - Informatique, Mathématiques et Applications de Grenoble (TIMC-IMAG), Centre National de la Recherche Scientifique (CNRS), Université Joseph Fourier (Grenoble 1), France

3. Centre National de la Recherche Scientifique (CNRS) and Université Pierre Mendès France (Grenoble 2), France

*To whom correspondence should be addressed. E-mail: edouard.gentaz@upmf-grenoble.fr

**Abstract**

**Background:** Learning to perform new movements is usually achieved by following visual demonstrations. Haptic guidance by a force feedback device is a recent and original technology which provides additional proprioceptive cues during visuo-motor learning tasks. The effects of two types of haptic guidances - control in position (HGP) or in force (HGF) – on visuo-manual tracking ("following") of trajectories are still under debate.

**Methodology/Principals Findings:** Three training techniques of haptic guidance (HGP, HGF or control condition, NHG, without haptic guidance) were evaluated in two experiments. Movements produced by adults were assessed in terms of shapes (dynamic time warping) and kinematics criteria (number of velocity peaks and mean velocity) before and after the training sessions. Trajectories consisted of two Arabic and two Japanese-inspired letters in Experiment 1 and ellipses in Experiment 2. We observed that the use of HGF globally improves the fluency of the visuo-manual tracking of trajectories while no significant improvement was found for HGP or NHG.

**Conclusion/Significance**: These results show that the addition of haptic information, probably encoded in force coordinates, play a crucial role on the visuo-manual tracking of new trajectories.

**Introduction**

Learning to perform new movements is usually achieved by following visual demonstrations [1]. Haptic guidance by a force feedback device is a recent and original technology that provides additional proprioceptive cues during visuo-motor learning tasks. Virtual simulators, in which haptic and visual cues are provided, seem to be an efficient way to teach complex movements [2-5]. Two well-known robotic haptic guidances have been currently implemented: The first one uses spatial coordinates (HGP) - position information - of the trajectory to learn, whereas the second one (HGF) uses forces generated by a teacher to control the student's task (Figure 1).

Haptic guidance in position (HGP) mostly uses a proportional derivative controller *i.e.* following point-per-point the visual representation of the target trajectory. Based on this technology, Solis *et al.* [6] had developed a Japanese calligraphy system using reactive robot technology. Unfortunately, this study mainly focused on the technical aspects. In the same vein, Henmi *et al.* [7] also designed a Japanese calligraphy system using a "record and playback" strategy: The authors recorded positions and forces applied by a human teacher and displayed them to the students. However, in these two studies, no behavioral data was reported. Gillespie *et al.* [8] developed a virtual teacher based on a proportional derivative position controller to help students to properly move a simulated crane. This pilot study with 24 participants showed that their implementation of the virtual teacher concept did not significantly improve the learning of oscillating curves. More recently, Palluel-Germain *et al.* [5], in a pilot study, analyzed the effects of using HPG to train the fluency of writing cursive letters in kindergarten children. Fluency of handwriting (analyzed by kinematics parameters such as average velocity, number of velocity peaks, and number of breaks during the production) was tested before and after the training sessions (either visuo-haptic or control). Letters were computer generated to control the dynamics by changing the distance between successive points of a discrete trajectory. Results showed that the fluency of handwriting for all letters was higher after the visuo-haptic training session than after the control training session: The movements of the hand were faster, exhibited less velocity peaks and the children lifted the pen less frequently during handwriting. Finally, other studies in adults [2, 4] confirmed the positive effect of visuo-haptic training sessions using proportional derivative position controller but most of them mainly describe the technical aspects. In these studies, the analysis of kinematics criteria remained rather unexplored.

Haptic guidance in force (HGF) is an alternative control method, which is congruent with two well-known psychophysical principles [9, 10, 11]: The homothety principle which states that the trajectory keeps its shape characteristics whatever its size, and the isochrony principle which states that the velocity for tracing increases as function of its size. Hemni *et al.* [7] compared the effects of HGF and HGP on a handwriting task in addition to visual cues. Preliminary results showed both techniques to be equally effective. Srimathveeravalli *et al.* [12] introduced a new paradigm providing the closest possible replication of an expert's skill. The authors proposed that if the nature of forces generated by the teacher and by the student were the same, then their trajectories would be similar. Force profiles of the teacher were then used to guide the motion of the student. Demonstration of its efficiency was shown by comparing this method with other classical haptic training methods in terms of shape matching with an unfamiliar Tamil (Indian) letter. Results confirmed the authors' hypothesis and showed that a "record-and-playback" training method with force information was more efficient than training method with only position control. Unfortunately, this study mainly focused on a shape matching score and did not examine kinematics criteria. Recently, Morris et al [13] explored the use of haptic feedback for teaching a sequence of forces. Results showed that adults are able to learn sensory-motor skills via visuo-haptic training. This result would allow us to better understand the positive effects of HGF during training session of handwriting observed by Srimathveeravalli [12].

In the present study, we have investigated with adults whether the two types of haptic guidance - control in position (HGP) or in force (HGF) – based on psychophysics laws of movement production, would improve visuo-manual tracking of Arabic and Japanese-inspired letters (Experiment 1) and untrained ellipses (Experiment 2). The effects of HGP on kinematics would then be tested, in extension of Palluel-Germain *et al.* [5] study. Moreover, the effects of HGF on kinematics criteria (fluidity) were considered as a complement of Srimathveeravalli [12] study. In both experiments three training sessions were conducted, which differed according to the haptic guidance used: HGP, HGF or no haptic guidance (NHG). We proposed that the addition of haptic cues to training session would improve the performance of subjects. Movements were evaluated in terms of shapes (dynamic time warping) and kinematics criteria (number of velocity peaks and mean velocity). Progress was assessed from the difference of performances before (pre-test) and after (post-test) the training session. We hypothesized that haptic guidance of both types would improve the performance of the subjects in comparison to the control training session.

**Experiment 1: Japanese and Arabic letters**

The goal of Experiment 1 was to explore whether the two types of haptic guidance - control in position (HGP) or in force (HGF) – improved the visuo-manual tracking of unfamiliar letters in terms of shape and kinematics aspects of handwriting. To test this goal, 23 adults were asked to learn to track visuo-manually two unfamiliar Arabic and two unfamiliar Japanese-inspired letters.

**Method**

Participants

Participants were 23 right-handed Caucasian adults, with no significant language, motor or neurological dysfunction. They were students from University of Grenoble and their age ranged from 18 to 26 years. The present study was conducted in accordance with the Declaration of Helsinki. It was conducted with the understanding and the written consent of each participant which was obtained and was approved by the local ethic committee.

Experimental Setup

The present experimental setup was similar to the "WYSIWYF" interface proposed by Yokokohji *et al.* [14]. We used a PHANToM™ Omni device (Sensable Technology). The modified PHANToM's stylus (Figure 2.a) served as a pen and a simple flat screen, mounted under the force feedback device, served as a paper. Figure 2.c shows a user writing with the virtual interface. The Chai3D Framework [15] was used to develop the application (Figure 2.b), on a classical personal computer (Pentium IV, 3.2 Ghz, 2Go Ram, NVIDIA Quadro Fx). Efforts have been made in the design of the physical setup to put the user in a situation, as close as possible to the usual handwriting task. The maximal depth from the virtual drawing and the real PHANToM's stylus is 0.5mm (height of the protective glass on the screen). Calibration between the force feedback device and the screen was done by triangulation at the beginning of each experiment. This resulted in co-location of the stylus and the virtual trace which provided a natural feeling of handwriting.

Tested and Trained trajectories: Arabic and Japanese-like letters

Trajectories were chosen to be "biologically and culturally possible" but also unfamiliar for participants (criterion used to select the participants). We explored two Arabic letters (fig. 3.a and 3.b) and two Japanese-inspired letters (fig. 3.c and 3.d). These trajectories were generated from several expert productions. Moreover, they were chosen to provide different difficulty levels defined by the number of "brutal change" of direction (> 45°) in the shape (letter 1: one 90° change; letter 2: one 180° change; letter 3: three changes; letter 4: four changes). It should be noted that theses changes of direction in shape imply large changes in the kinematics. The Japanese letters were modified in order to avoid lift up from the stylus by orthographically projecting the aerial path in the 2-dimensional reference of the letter.

Pre-test and post tests

Participants were assessed before and after the training sessions in order to measure the visuo-manual tracking of letters. Participants were asked to trace with their right hand visually presented letters with the stylus as accurately and as promptly as possible. No feedback was given by the experimenter. Each participant executed five trials for each letter in pseudo random order (two identical consecutive letters never occurred). In total, there were therefore 20 trials in the pre and post-tests.

Movement analysis

Three main criteria were used to evaluate the movement: Number of velocity peaks, mean velocity and shape matching score. The positions and time stamps of each trial were recorded and pre-processed to compute these measures. The PHANToM device allowed us to record the position of the tool tip at about 1 KHz, which resulted in over sampling. To avoid long computation time and due to the frequency range of information, data could be reasonably sampled at 200Hz. This is equivalent to applying a low-pass filter to avoid high frequency noise due to hardware imprecision in time sampling.

*1. Number of Velocity Peaks*

The number of velocity peaks is a criterion to estimate the fluidity of the movement. A small number of velocity peaks, with no regression of the shape quality (shape matching criterion) indicates a "good" fluidity. To compute these peaks, the velocity was low-pass filtered using a 6-order Butterworth filter (cut-off frequency = 50Hz). Attention was paid to avoid distortions. Then, we computed the acceleration from these filtered data and counted the sign inversions for the acceleration. This value gave us the number of velocity peaks.

*2. Mean Velocity*

The mean velocity also is a criterion to estimate the fluidity of the movement. A high mean velocity, with no regression of the shape quality (shape matching criterion), indicates a good fluidity of handwriting production.

*3. Shape matching*

Dynamic time warping (DTW) algorithm [13, 16] computes a cost corresponding to a match between a reference trajectory and a subject recall trajectory. DTW constructs a global cost matrix by aligning the two temporal series. Then, a minimal path through the matrix is determined and the final value of this minimal path provides a representative cost for the warping of the two trajectories, *i.e.* the similarity between the two shapes. This algorithm was implemented for each axis (X and Y of the unit table) separately. The cost function in this case was the Euclidean distance between two points. This criterion gives a score of shape matching: A low score means a good match in shape.

Training Sessions

Each subjects participated in three training sessions (Haptic guidance in position - HGP, haptic guidance in force – HGF, and no haptic guidance – NHG). The order was given by a Latin-square plan. Participants were asked to move a stylus with their right hand to follow a visually presented trajectory. There were 20 test trials (5 trials x 4 letters) in each training session.

*1. Haptic Guidance in Position (HGP)*

Position Proportional-Derivative control is well known in automatism. It consists in minimising the trajectory error during the training. This type of control can be thought of as a spring, attached to the trajectory points, pulling the stylus' tip to the next point (Figure 1.a). The proportional and derivative gains were derived from [12] and experimentally tuned (with pilot studies) to 0.4 N/mm and 0.2 Ns/mm.

*2. Haptic Guidance in Force (HGF)*

According to the hypothesis proposed by Srimathveeravalli *et al.* [12], similar force profiles lead to similar trajectories (Figure 1.b). The handwriting task has been modelled as a force needed to move a mass on a surface with a constant friction value. To compute the forces from computer generated trajectories, a simple lumped model, described in previous studies [12], was used. This model is described in Equation (1).

$$F = m.a + c.v + k.x \qquad (1)$$

where $F$ is the force, $m$ is the mass of the system (hand + stylus), $a$ is the acceleration, $c$ is the damping coefficient and $k$ is the spring constant. According to this simple model, we were able to compute the force profile for the previously generated trajectories. The mass, the damping coefficient and the spring constant used for the method were theoretically estimated and were set equal to 0.1Kg, 0.5Ns/mm and 0.1Ns/mm respectively. This force profile was then used as a reference for a proportional derivative control with visual tracking. Even if visual matching errors could occur during the movement, the haptic sensations felt by the user would be similar to what the expert felt during his interaction with the model. The proportional and derivative gains used were theoretically estimated and were set to 0.5N/mm and 0.1Ns/mm respectively.

*3. Control session without haptic guidance (NHG)*

To replicate learning through observation and manual repetition, no haptic assistance was provided to the participants during the training session. This task was similar to pre and post tests in which no haptic guidance was added.

**Results**

Preliminary analysis of variance (ANOVA) showed that the order of training sessions had no effect and did not interact with any other factors (all p> .25). Then, for each criterion, ANOVA was performed with test (pre and post tests), and letter (L1: Arabic Letter 1, L2: Arabic Letter 2, L3: Japanese-inspired Letter 3 or L4: Japanese-inspired Letter 4) as within factors and training mode (HGF, HGP or NHG) as independent factors. Summary of raw data can be found in Table 1.

*1. Number of velocity peaks*

The main effect of letters was significant (F(3,66)=11.20; p<.05). Pre-planned contrasts showed (all p<.05) that the number of velocity peaks increased with the type of letters: Letter 1 (m=8.77), Letter 2 (m=9.27), Letter 3 (m=11.03) and Letter 4 (m=10.02). This factor did not interact with training mode. The interaction between training session and period factors was significant (F(2,44)=5,01; p<.05). Post-hoc analyses (Tukey test; p<.5) revealed a significant decrease of the number of velocity peaks for only HGF mode. However, no significant difference was observed for both HGP and NHG modes. Finally, the interaction between training mode and letter factors was not significant (F(6,102)=1,15; p>.25).

We have performed additional analysis in order to asses the effects of training mode on each test of letter type:

-Arabic Letter 1 and Japanese-inspired Letter 3: No significant interaction between training mode and test.

-Arabic Letter 2: The interaction between training mode and test was significant (F(2,44) = 6.14; p<.05). Post-hoc analyses (Tukey test; p< .05) revealed a significant decrease for HGF (Pre-test: m=9.42; post-test m=7.06) mode and no significant difference for NHG and HGP modes.

-Japanese-inspired Letter 4: The interaction between training mode and test was significant (F(2,44) = 3.79; p<.05). Post-hoc analyses (Tukey test; p< .05) revealed a significant decrease for HGF (Pre-test: m=11.12; post-test m=8.86) mode (p<.05) and no significant differences for NHG and HGP modes.

*2. Mean velocity*

The main effect of letters was significant (F(2,66)=13.577; p<.05). This factor did not interact with training mode. The interaction between training mode and test was significant (F(2,44)=11.09; p<.05). Post-hoc analyses (Tukey; p<.05) revealed a significant increase of the mean velocity for only HGF mode. By contrast, no significant difference was observed for both HGP and NHG mode.

We have performed additional analysis in order to asses the effects of training mode on each test of letter type:

-Arabic Letter 1: The interaction between training mode and test was significant (F(2,44)=8.734;p<.05). Post-hoc analyses (Tukey; p<.05) revealed a significant decrease for HGF mode

-Arabic Letter 2: The interaction between training session and period factors was significant (F(2,44)=13.135;p<.05) Post-hoc analyses (Tukey test; p<.05) revealed a significant increase for HGF

-Japanese-inspired Letter 3: The interaction between training mode and test was significant (F(2,44)=11.559;p<.05). Post-hoc analyses (Tukey test; p<.05) revealed a significant increase for HGF mode and no significant differences for NHG and HGP modes

-Japanese-inspired Letter 4: The interaction between training mode and test was significant (F(2,44)=7.744;p<.05). Post-hoc analyses (Tukey test; p<.05) revealed a significant increase for HGF mode and no significant differences for NHG and HGP modes.

*3. Shape Matching Score*

The main effect of letters was significant (F(3,66)=277.01; p<.05). This factor did not interact with training mode factor. The interaction between training mode and test was not significant (F(1,22)=0.61; p>.25). Complementary analyses per letters were then performed to precise the specific effect of training mode on each letter after training sessions and no significant effect was observed (all p>.25).

**Discussion**

The main results of this experiment revealed a significant reduction of the number of velocity peaks (for two among the four letters) and a significant increase on mean velocity through all the tested letters with HGF training. The results are concordant with our hypotheses, even though this type of haptic guidance appeared to be sensitive to the trajectories tested. Contrary to our hypotheses, no major effect on the number of velocity peaks and the mean velocity was found for HGP. It appears that HGP has no significantly beneficial effect on the fluidity of movements (even if trends of improvement were observed). No effect of NHG training was found on all criteria. Finally, none of the training modes effected shape matching criterion. Differences between our and previous results from literature could be explained by the different types of trajectories used. Moreover, it is possible that same trajectories for the training and the test could have weakened the power of the results. To test this hypothesis, we have investigated this relation in a more detailed experiment with specifically chosen trajectories. Moreover, we added variability in trajectories by changing the experimental protocol (tested and trained trajectories were not the same).

**Experiment 2: Untrained ellipses**

The goal of the second experiment was to explore whether training on one set of trajectories with the two haptic guidances - HGP or HGF – improve the visuo-manual tracking of another set of similar trajectories in terms of shape and kinematics aspects. In Experiment 1, HGP did not show any overall improvement on performance. We have proposed shape variability of stimuli to improve the training sessions. Generation of a range of trajectories required a highly defined base trajectory. The ellipse, a "two-parameter" trajectory (well-known in psychophysics studies) was chosen. Twenty-four adults were asked to track three visually presented ellipses. Psychophysics studies showed that there is an unambiguous relationship between letter production and movement kinematics. The shape of the trajectory determines the movement kinematics (so called "the two-thirds power law") [9, 10, 11]. The trajectory and dynamics of the drawing movement are mutually constrained by this law. Viviani *et al.* [9] found that this law can be taken as an explanation of steady-state adult performances of handwriting skills. The "two-third" law was used to generate trajectories for this Experiment in order to prevent any dependence of the teacher's specific way of tracing a trajectory. We introduced shape variability in the trained trajectories. They were similar to the pre and post test trajectories but never the same. Because this variability of required movements with haptic guidances generated a variability in the sensorial feedbacks (visual and proprioceptive), we hypothesized an increase of performances to track new (but closed) movements.

**Method**

Participants

Participants were 24 right-handed adults, with no significant language, motor or neurological dysfunction. They were students from University in Grenoble and their age ranged from 18 to 28 years. The present study was conducted in accordance with the Declaration of Helsinki. It was conducted with the understanding and the written consent of each participant which was obtained and was approved by the local ethic committee.

Experimental Setup

The experimental setup was the same as in Experiment 1.

Pre-test and post tests

Participants were assessed before and after the training sessions in order to measure the visuo-manual tracking of letters. Participants were asked to trace with their right hand visually presented ellipses with the stylus as accurately and as promptly as possible. No feedback was given by the experimenter. Each participant followed 18 ellipses presented in pseudo random order. In total, there were therefore 18 trials in the pre test and 18 trials in the post-test.

Characteristics of test Trajectories

In the pre and post tests, three target trajectories were derived from a base ellipse. Contrary to the Record-and-playback strategy [12], where the base trajectories were recorded from a teacher, we have designed elliptical paths by controlling each parameter (size, velocity profile, number of points...). The shape of each ellipse varied from 6 to 2 cm in width and in height. The generation of these trajectories was made by a Scilab™ (www.scilab.org) routine. Stimuli used for this experiment were composed of 1000 points (X,Y). Their velocity profiles followed the two-third power law [10, 11], (*i.e.*, velocity V is proportional to the radius of curvature r of the trajectory: $v = k * r^{-1/3}$; equivalent to angular velocity A is proportional to the curvature c of the trajectory: $A = k * c^{2/3}$). Our generated trajectories were in adequacy with a trained-level movement skill. The untrained tested trajectories are shown (red curves) in Figure 4.

Movement analysis

Three main criteria (number of velocity peaks, mean velocity and Shape matching) used to analyse the movements were the same as in Experiment 1.

Training Sessions

Three training sessions - HGP, HGF and NHG - were proposed to each participants (the order was given by a Latin-square plan). There were 24 test trials in each training session. The trajectories used during the pre and post test were never encountered during the training sessions. This was done to provide variability during the training session. Ellipses used in training were generated using the same procedure as trajectories used in tests (cf. § Characteristics of Test Trajectories). Eighteen ellipses have been chosen around the three tested ones (see Figure 4). They appeared randomly during the training session. In total, each participant performed 72 trials. The three haptic guidance modes and parameters were similar to those found in Experiment 1 except for HGP, where the proportional and derivative gains were experimentally tuned to 0.6 N/mm and 0.2 Ns/mm for smoother sensations.

**Results**

Preliminary analysis of variance (ANOVA) showed that the order of training sessions had no effect and did not interact with any other factors (all p> .25). Then, for each criterion, ANOVA was performed with test (pre and post tests), and ellipse (E1, E2 or E3) as within factors and training mode (HGF, HGP or NHG) as independent factors. Summary of raw data can be found in Table 2.

*1. Number of velocity peaks*

The main effect of tested ellipses was significant ($F(2,46)=16.342$; $p<.05$). Pre-planned contrasts showed that the number of velocity peaks was lower for the circle (E2: m=9.3) than the horizontal (E1: m =13.2) and vertical (E2: m=14.1) ellipses ($F(1,23)=19.67$; $p<.01$) This factor did not interact with training mode factor. The interaction between training mode and test was significant ($F(2,46)=8.86$; $p<.05$). Post-hoc analyses (Tukey test; $p<.05$) revealed a significant decrease of the number of velocity peaks for both HGP and HGF mode. By contrast, no significant difference was observed for NHG mode.

*2. Mean velocity*

The main effect of tested ellipses was significant ($F(2,46)=19.321$; $p<.05$). Pre-planned contrasts showed that the mean velocity was lower for the circle (E2: m=4.70 cm/s) than the horizontal (E1: m=5.86 cm/s) and vertical (E2: m=5.77 cm/s) ellipses ($F(1,23)=30.53$; $p<.05$) This factor did not interact with training session factor. The interaction between training mode and test was significant ($F(2,46)=13.22$; $p<.05$). Post-hoc analyses (Tukey test; $p<.05$) revealed a significant increase of mean velocity for only HGF mode. However, no significant difference was observed for both HGP mode and NHG mode.

*3. Shape Matching Score*

The main effect of tested ellipses was significant ($F(2,46)=12.482$; $p<.01$). Pre-planned contrasts showed that the mean DTW shape matching score was lower for horizontal ellipse (E1: m=20.13) and the circle (E2: m=20.18) than vertical (E2: m=27.53) ellipses ($p<.05$). This factor did not interact with training mode. The interaction between training mode and test was not significant ($F(2,46)=0.45$; $p>.25$).

**General Discussion**

The present study examined whether two well-known types of haptic guidance - HGP or HGF - improve the visuo-manual tracking of trajectories. The number of velocity peaks and mean velocity were the two criteria used to estimate the fluidity of movements. The results of Experiment 1, in which trained and tested movements were identical, showed that HGF mode reduced the number of velocity peaks for only two among four letters and increased mean velocity. These results were consistent with our hypotheses and within previous literature [12, 13]. The other two modes, HGP and NHG, had no significant effect on these two performances criteria. Detailed analysis revealed similar effects of training modes, independent of the type of the letter: HGF improved performances whereas HGP and NHG showed no significant improvement. The lack of effect of HGP was not consistent with our hypotheses and results observed by Feygin *et al.* [2] or Teo *et al.* [4]. The efficiency of haptic guidances with different difficulty levels of trajectories could be discussed in relation to the choice of parameters (proportional and derivative gains) because their respective influence is not clearly established.

In the second experiment, results showed that both HGF and HGP reduced the number of peaks during visuo-manual tracking of the test ellipses. These results were in concordance with our hypotheses and extended the results of Palluel-Germain *et al.* [5] with children. In the control group (NHG), it seems that visual feedback alone was not enough to improve the performance. However, only HGF mode increased the mean velocity on test ellipses in contrast to NHG and HGP modes. This showed that HGF better improved the fluidity of movements than HGP. Contrary to our results, Palluel-Germain *et al.* [5] observed that HGP increased mean velocity with children. This suggests that HGP may be less suitable for adults since adults would have better knowledge of the shape they had to draw and better kinesthetic control of their upper limbs. The global superiority of HGF over HGP suggested that learned information for this specific motor activity could be stored as internal inverse model and encoded in force coordinates as suggested by Krakauer [17]. This suggests taking into account the type of feedback information included in these internal models because the effects of visual, position or force information were non equivalent. Thus, kinematics information could be encoded in reference to force coordinates rather than spatial Cartesian coordinates. We could also discuss these results with respect to different internal representation stages of motor knowledge, which evolve from children to adults (creation of internal representation - internal inverse models [18] or generalized motor plan - or adjustment of this knowledge). Moreover, our results could be explained by the generalization of shapes during training session that helped to integrate these trajectories. By providing several training trajectories, adjustment of internal representations could be more involved and thereby improved. This proposes that a motor learning task (as drawing) would be improved by variability of required movements as evidenced in sports [19]. Further questions remain unanswered: As human handwriting production is variable by nature, would a handwriting training with variability in letters be efficient? In which way can we introduce variability in letters? Could variability within the letters be a set of several letters produced by experts? Further theoretical definition of variability needs to be investigated.

Finally, Dynamic Time Warping (DTW) score gives a score (level) of shape matching between two trajectories: Expected and recorded trajectories. This criterion was used in previous studies [12] to assess the effect of haptic training on the shape of a trajectory. In both experiments, no effect of training modes on the shape matching criterion was found. These former results did not confirm the effect on shape matching observed by Srimathveeravalli et al [12]. The lack of improvement could be explained by a "ceiling effect" on the shape matching criterion, due to expert ability of adults and by different experimental designs (the authors [12] hide the model trajectory during recall phase, thereby suppressing any visual help).

In conclusion, we found a positive global effect of the HGF mode on the fluidity of movements in the two experiments. The superiority of this type of haptic guidance suggests that position information is mainly given by visual modality (no improvement of shape by adding haptic guidance) and kinematics information is given by haptic modality, probably encoded in force coordinates. Moreover, this study explored the use of variability in learning sessions: knowledge extracted from a set of trajectories (elliptical paths) during the training period of haptic guidance can be applied to unfamiliar trajectories of the same type, suggesting a generalization process.

**Acknowledgments**

We thank the student of psychology of University Pierre Mendès France who participated in the experimental study. We also thank Sophie Zijp-Rouzier and Nicolas Tarrin for their fruitful collaboration. Finally, we sincerely thank Jean-françois Cuniberto for his useful help building the physical setup and Richard Palluel-Germain, David Meary, Sylvette Maniguet and Inna Tsirlin for their help.

**Figure Legends**

**Figure 1** – Schematic view of haptic guidances: (a) Haptic guidance in position (HGP); the force felt by the user at time *t* is proportional to displacement between the current user position and the theoretical position on the model trajectory; (b) Haptic guidance in force (HGF); the force felt by the user at time *t* is the same as the force existing for the theoretical trajectory at the same time.

**Figure 2** – System overview: (a) The modified stylus pen; (b) The graphic User Interface displayed to the subject; (c) A subject undergoing training on the WYSIWYF interface.

**Figure 3** – Letters proposed in experiment 1: Letters 1 and 2 are Arabic and letters 3 and 4 are "Japanese-like" letters.

**Figure 4** – All ellipses used in experiment 2: In red, the three references trajectories used before and after each training session; In green and blue, the trajectories used during the training sessions, equidistant in the choice of their diagonals (eccentricity).

**Table 1** – Summary of raw data of Experiment 1 (mean ± SE).

Note: * means a significant difference given (post-hoc Tukey test; <.05).

|  | Number of velocity peaks |  | Sig. | Mean Velocity (cm/s) |  | Sig. | DTW |  | Sig. |
| --- | --- | --- | --- | --- | --- | --- | --- | --- | --- |
|  | Pre-test M ± SE | Post-test M ± SE |  | Pre-test M ± SE | Post-test M ± SE |  | Pre-test M ± SE | Post-test M ± SE |  |
| NHG | 9.87 ± 1.43 | 9.99 ± 1.42 |  | 5.8 ± 0.5 | 5.6 ± 0.45 |  | 45.77 ± 2.21 | 45.09 ± 1.57 |  |
| HGP | 10.17 ± 1.37 | 9.21 ± 1.05 |  | 5.40 ± 0.43 | 5.69 ± 0.43 |  | 45.38 ± 2.34 | 46.39 ± 2.6 |  |
| HGF | 10.82 ± 1.16 | 8.58 ± 1.16 | * | 4.97 ± 0.4 | 6.34 ± 0.52 | * | 45.29 ± 1.83 | 45.90 ± 2.14 |  |

**Table 2** – Summary of raw data of Experiment 2 (mean ± SE).

Note: * means a significant difference given (post-hoc Tukey test; <.05).

|     | Number of velocity peaks | | Sig. | Mean Velocity | | Sig. | DTW | | Sig. |
| --- | --- | --- | --- | --- | --- | --- | --- | --- | --- |
|     | Pre-test M ± SE | Post-test M ± SE | | Pre-test M ± SE | Post-test M ± SE | | Pre-test M ± SE | Post-test M ± SE | |
| NHG | 11.36 ± 1.89 | 13.80 ± 2.37 | | 5.66 ± 0.55 | 5.14 ± 0.61 | | 23.85 ± 2.77 | 21.18 ± 1.75 | |
| HGP | 14.48 ± 2.37 | 10.19 ± 1.58 | * | 5.13 ± 0.63 | 5.88 ± 0.52 | | 20.83 ± 1.48 | 20.39 ± 1.23 | |
| HGF | 14.19 ± 1.91 | 9.20 ± 1.38 | * | 4.62 ± 0.48 | 6.23 ± 0.54 | * | 22.84 ± 2.9 | 21.99 ± 1.43 | |

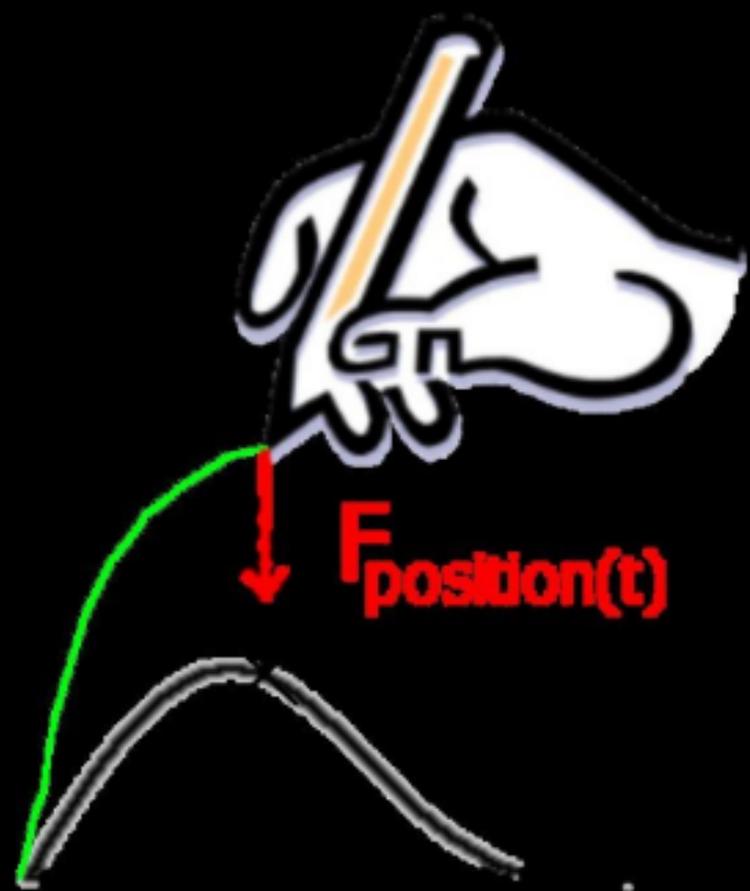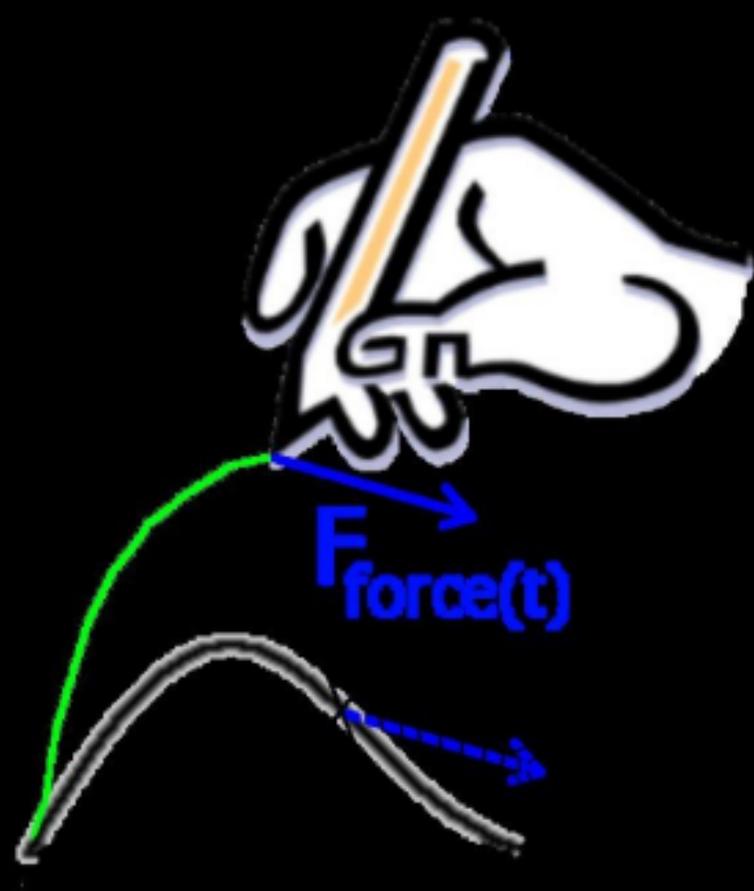

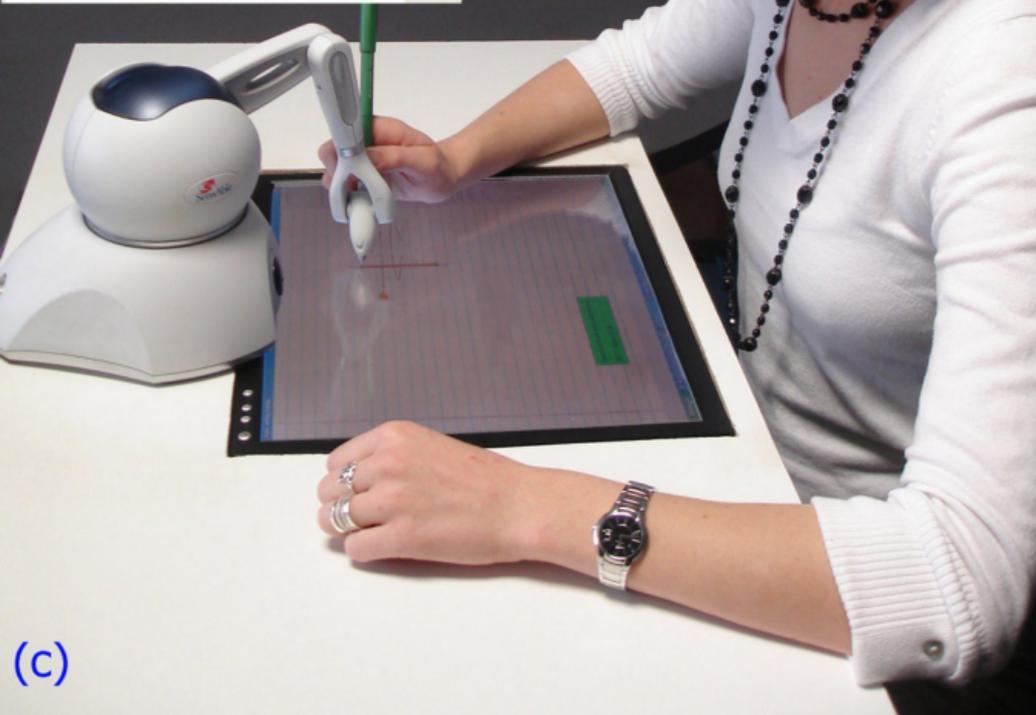

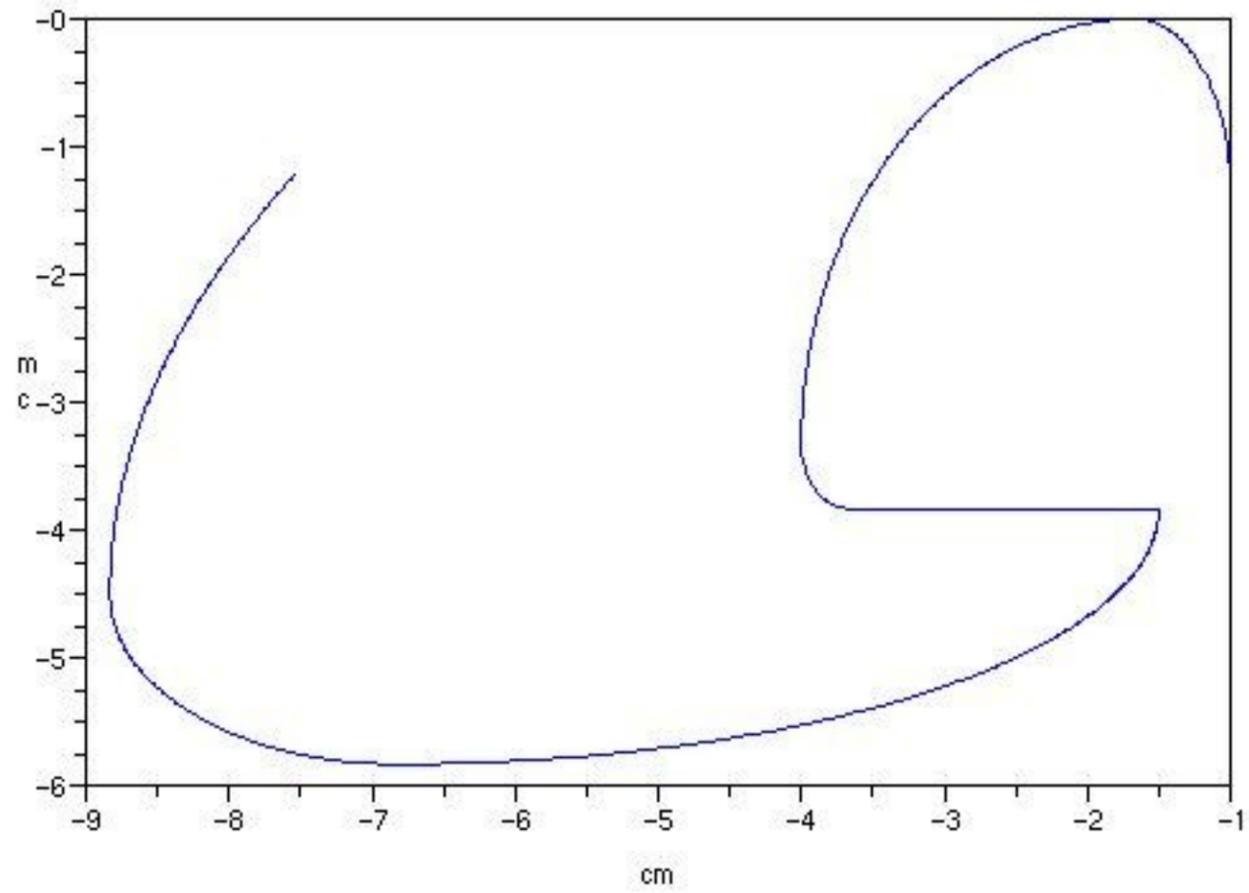

Arabic Letter #1

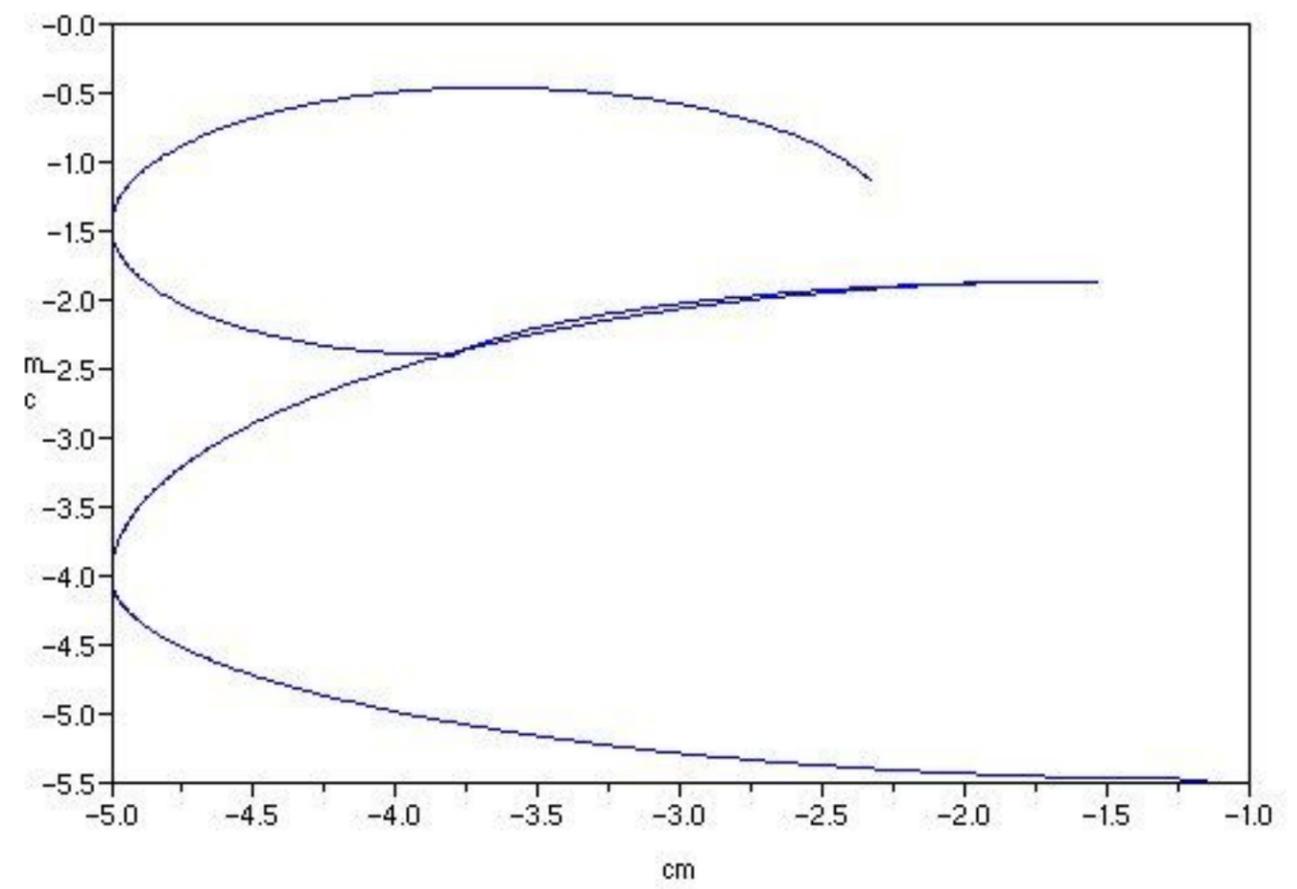

Arabic Letter #2

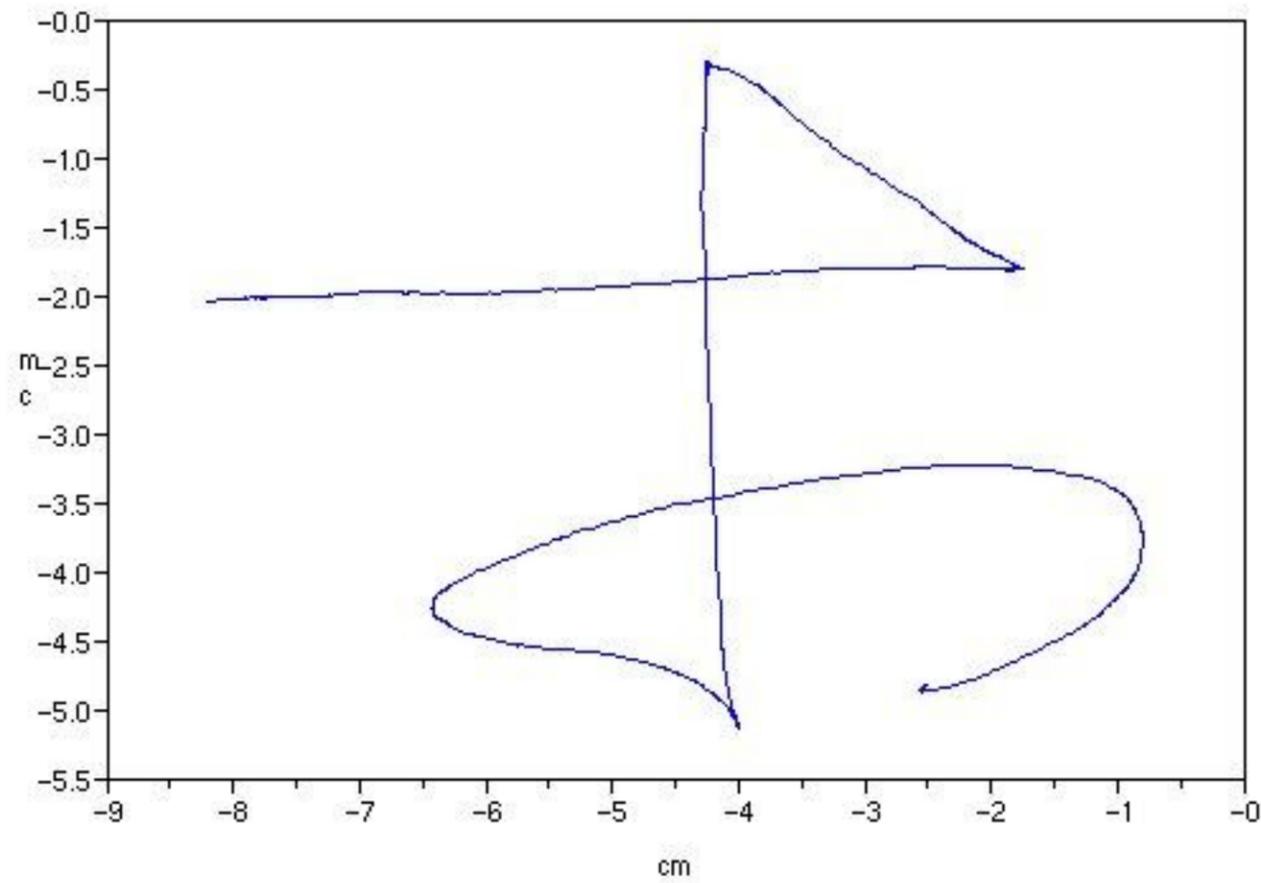

Japanese inspired Letter #3

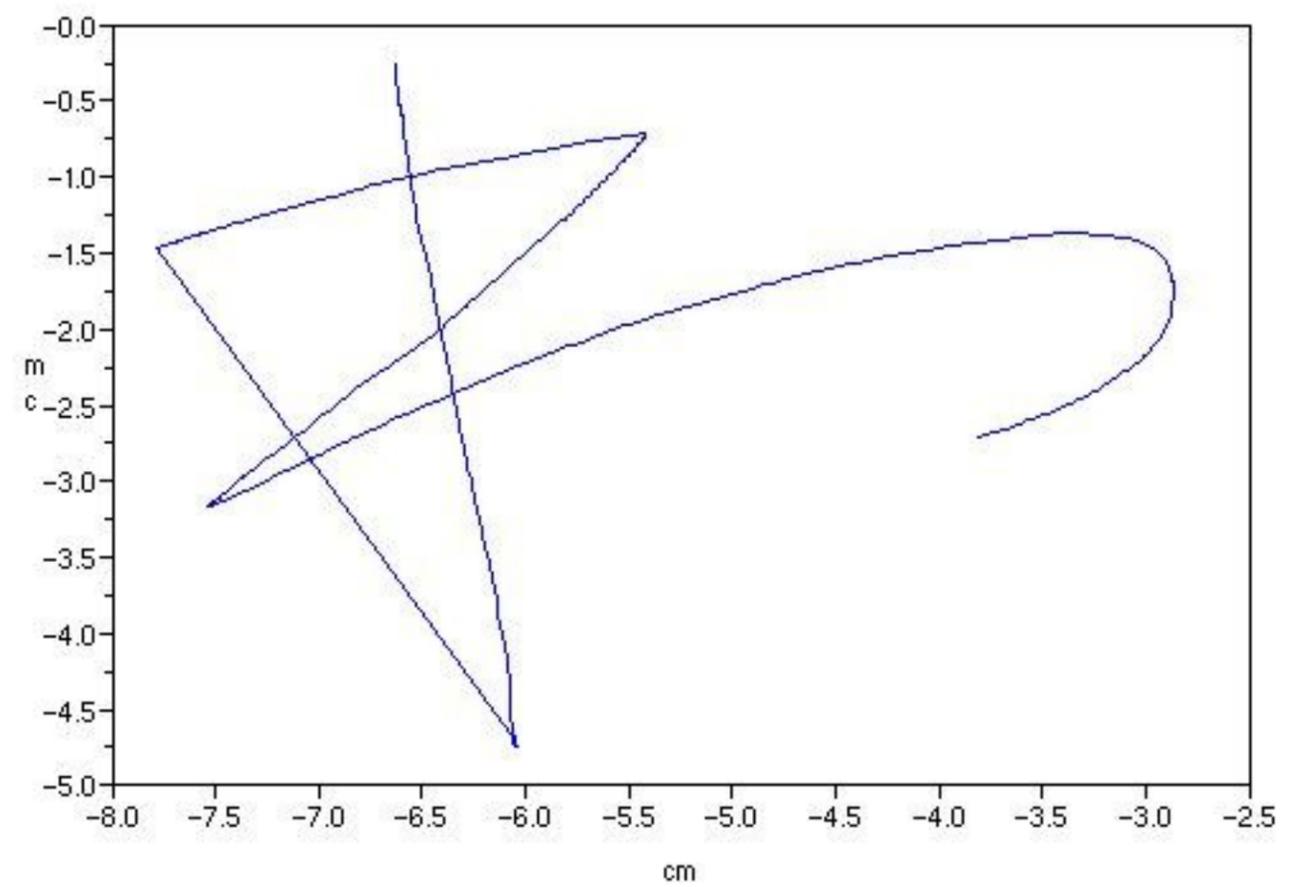

Japanese inspired Letter #4

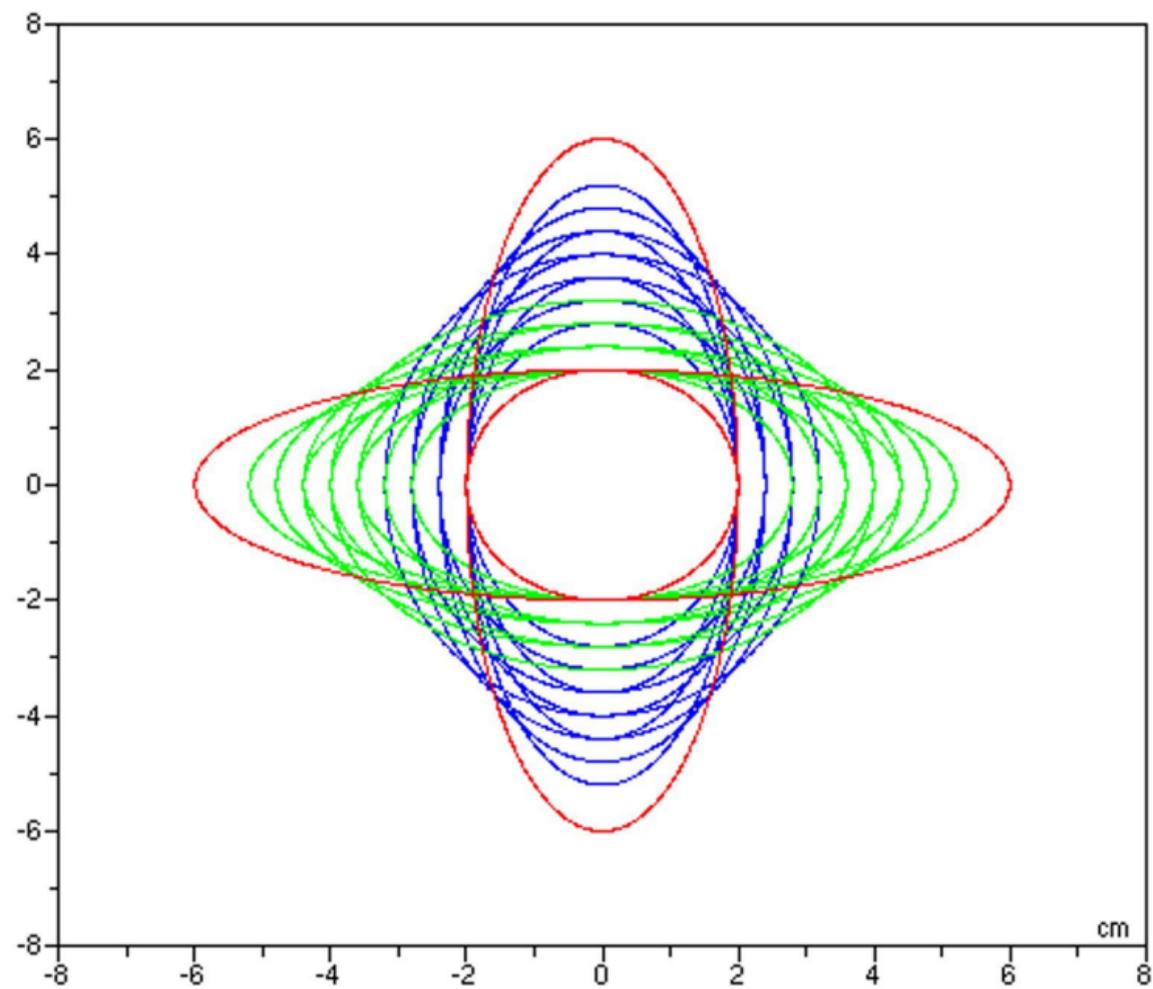